\documentclass[fleqn,12pt,twoside]{article}
\usepackage{amsmath}
\usepackage{espcrc1}
\usepackage{graphicx}

\newcommand{\be}{\begin{equation}}
\newcommand{\ee}{\end{equation}}
\newcommand{\bea}{\begin{eqnarray}}
\newcommand{\eea}{\end{eqnarray}}
\newcommand{\bvec}[1]{\mbox{\boldmath $#1$}}

\title{Casimir scaling hypothesis on the 
nonperturbative force in QCD\\
vs. dual superconducting scenario of 
confinement\footnote{
Talk given by Y. Koma at ``PaNic02'', Osaka, Japan, Sep.30 - Oct.4, 2002}
}

\author{Y. Koma\address[MPI]{
Max-Planck-Institut f\"ur Physik, 
F\"ohringer Ring 6, D-80805
M\"unchen, Germany}, 
M. Koma\addressmark[MPI], 
and 
H.~Toki\address[RCNP]{
Research Center for Nuclear Physics, 
Osaka University, 
Osaka 567-0047,
Japan}}  

\begin{document}

\maketitle

\begin{abstract}
We discuss the Casimir scaling hypothesis on the
nonperturbative force in terms of the dual superconducting
picture of the QCD vacuum 
by calculating the string tensions of flux tubes
associated with static charges in various SU(3) 
representations in the dual Ginzburg-Landau (DGL) theory.
\end{abstract}

\parskip=0.1cm

\vspace*{-0.3cm}
\section{Introduction}
\vspace*{-0.3cm}

\par
From the SU(2) and SU(3) lattice QCD studies of 
the static potential between 
color charges in various dimensional representations,
there arises the Casimir scaling hypothesis 
for the intermediate distance 
force~\cite{Bernard:1983my,Deldar:1999vi,Bali:2000un}.
This hypothesis tells that the ratio of forces 
associated with various dimensional 
representations of color charges, for a given 
color group, is determined 
by that of eigenvalues of the quadratic  
Casimir operator for each representation.



\par
It is expected that such a hypothesis
is realized for the short distance
force as described by one-gluon exchange,
since the coupling is proportional to the quadratic 
Casimir operator.
However, it is hard to imagine that 
this property is kept until intermediate distance 
where the nonperturbative effects set in.
If the behavior of the ratio is governed exclusively 
by the group theoretical factor it should be 
manifest in arbitrary  SU($N$) gauge theory.
Recent studies of $k$-strings in SU(4) 
and SU(6) lattice gauge theories, however, 
do not support Casimir 
scaling~\cite{DelDebbio:2001kz,Lucini:2001nv}.
Therefore, it seems natural to consider that 
there are nonperturbative dynamics, 
rather than Casimir factor, which has inspired
the Casimir scaling hypothesis for the intermediate 
distance force in early lattice investigations.


\par
In this paper, we show that the dual superconducting 
picture of the nonperturbative QCD vacuum, as practically 
described by the dual Ginzburg-Landau (DGL) theory, 
provides us an understandable idea to explain the 
mechanism hidden behind the lattice data.
In the dual superconducting vacuum, the color-electric 
flux is squeezed almost one dimensional tube, called
the flux tube, due to the dual Meissner effect.
For the quark and the antiquark system, then
the flux tube is formed between the sources,
leading to the linear potential, where its
slope is identified as the  string tension.

\vspace*{-0.3cm}
\section{Flux-tube solution in the DGL theory}
\vspace*{-0.3cm}


\par
We examine the SU(3) case and calculate 
the string tensions of flux tubes 
in the DGL theory associated 
with static charges in various SU(3) representations
systematically~\cite{Koma:2001ut,Koma:2001is}, 
starting from the manifestly Weyl symmetric 
form~\cite{Koma:2000wn,Koma:2000hw}
\bea
&&
{\cal L}_{\rm DGL} 
= 
\sum_{i=1}^3 \Biggl [ 
- \frac{1}{6g^{2}}{}^{*\!} F_{i\;\mu\nu}^{2} 
+ \left | \left 
(\partial_{\mu} + i B_{i\;\mu}\right )\chi_{i} \right |^2 
- \lambda \left ( \left |\chi_{i} \right |^2-v^2 \right )^2 
\Biggr ] ,\\*
&&{}^{*\!}F_{i\; \mu\nu} = 
\partial_{\mu}B_{i\;\nu} - \partial_{\nu}B_{i\;\mu}
+2\pi \sum_{j=1}^{3} m_{ij}
\Sigma_{j\;\mu\nu}^{(e)}.
\label{eqn:dgl-weyl}
\eea
Here $B_{i\;\mu}$ and 
$\chi_{i}$  $(i=1,2,3)$
denote the dual gauge field and the complex scalar 
monopole field, respectively. 
The dual gauge fields within the Weyl symmetric 
expression have a constraint 
$\sum_{i=1}^{3}B_{i\;\mu}=0$.  
The $\Sigma_{j\;\mu\nu}^{(e)}$  $(j=1,2,3)$ 
in the dual field strength tensor
is the  color-electric Dirac string and 
their boundaries define the quark current 
$j_{j\;\mu}^{(e)}$ $(j=1,2,3)$.

\par
The color-electric charge of the quark
is specified by  the weight vector of the SU(3) 
algebra, $\vec{w}_{j}$
($j=1,2,3$), as $\vec{Q}_{j}^{(e)} 
\equiv e \vec{w}_{j}$,  where
$\vec{w}_1= \left (1/2, \sqrt{3}/6 \right )$, 
$\vec{w}_2= \left (-1/2, \sqrt{3}/6 \right )$, and 
$\vec{w}_3= \left (0, -1/\sqrt{3} \right )$.  
On the other hand, the color-magnetic charges of
the monopole fields $\chi_{i}$ are expressed by the root
vectors of the SU(3) algebra, 
$\vec{\epsilon}_i$, as 
$\vec{Q}_{i}^{(m)} \equiv g \vec{\epsilon}_i$
($i=1,2,3$), where 
$\vec{\epsilon}_1=\left (-1/2,\sqrt{3}/2 \right )$,
$\vec{\epsilon}_2=\left(-1/2,-\sqrt{3}/2 \right )$, 
and
$\vec{\epsilon}_3=\left (1,0 \right )$.  
These color-electric and 
color-magnetic charges satisfy
the extended Dirac quantization condition 
$\vec{Q}_{i}^{(m)} \cdot
\vec{Q}_{j}^{(e)} = 2\pi m_{ij}$, 
where $m_{ij} =2 \vec{\epsilon}_i
\cdot \vec{w}_{j}$ and $eg=4\pi$.
There are two mass scales in the DGL theory:
the masses of the dual gauge boson $m_B = \sqrt{3}gv$ and the 
the monopole field $m_\chi = 2\sqrt{\lambda}v$.  
In analogy to usual superconductors, 
their ratio, $\kappa \equiv m_\chi/m_B$,
corresponds to  the Ginzburg-Landau (GL) parameter,
which describes the type of dual superconductivity 
of the vacuum.


\par
The flux-tube solution is obtained by 
considering the cylindrical symmetric 
system with  translational invariance along the $z$ axis.
The fields are described as functions 
of radius $r$ and azimuthal angle $\varphi$.
Thus, we  write the modulus of the 
monopole field as $\phi_{i}(r) =|\chi_{i}(r)|$.
The dual gauge field is parametrized as
$\bvec{B}_{i} = 
[B_{i}^{\rm reg}(r) +B_{i}^{\rm sing}(r)]
\bvec{e}_{\varphi}$
where
$B_{i}^{\rm reg}(r) = [\tilde B_{i}^{\rm reg}(r)/r]$ 
and 
$B_{i}^{\rm sing}(r) = -n_{i}^{(m)}/r$ 
with
$n_i^{(m)} \equiv \sum_{j=1}^{3}m_{ij}n_{j}^{(e)}$.
Here $n_j^{(e)}$ is the winding number of $j$-type 
color-electric Dirac string $\Sigma_{j\;\mu\nu}^{(e)}$, 
which takes various integers depending on the 
representation of the SU(3) color group to which
the charges belong. 
For a given representation $D$ of the SU(3) group,
denoted by the Dynkin index $[p,q]$, we have
$\{ n_{1}^{(e)},n_{2}^{(e)},n_{3}^{(e)} \}
=\{ p, -q, 0 \}$~\cite{Koma:2002cv}.

\par
The string tension of the flux tube is calculated 
as an energy per unit length in $z$ direction:
\bea
\sigma_D 
=
2\pi  \sum_{i=1}^3   \int_0^{\infty}
\!\!\! 
rdr
\Biggl [
\frac{1}{3g^2} \left ( \frac{1}{r}\frac{d \tilde B^{\rm reg}_{i}}
{dr} \right )^2
\!\!  + \!    \left ( \frac{d \phi_i}{d r} \right)^2
\!\!   +\!
\left ( 
\frac{ \tilde B^{\rm reg}_{i} -  n_{i}^{(m)}}{r} 
\right )^2 \! \! \phi_i^2 + \lambda ( \phi_i^2 - v^2 )^2
\Biggr ] .
\label{eqn:string-tension}
\eea
In the Bogomol'nyi limit, $\kappa=m_\chi/m_B=1$, 
this expression is analytically
evaluated as~\cite{Koma:2001ut,Koma:2001is,Chernodub:1999xi}
\be
\sigma_D = 2 \pi v^2 \sum_{i=1}^3 \left | n_{i}^{(m)} \right | 
=    4 \pi v^{2} (p+q) .
\label{eqn:st-exact}
\ee
In this case, the ratio of the string tension between 
a higher and the fundamental representation 
$[1,0]$ is found to be 
$d_{D}= \sigma_{D}/\sigma_{\bf 3}=p+q$.
In the general dual superconducting vacuum 
of type~I ($\kappa<1$) and of type~II ($\kappa>1$),
one has to evaluate the whole 
expression~\eqref{eqn:string-tension}
in its variational minimum by solving 
the field equations numerically.


\begin{figure}[!t]
\begin{center}
\includegraphics[ width=11cm]{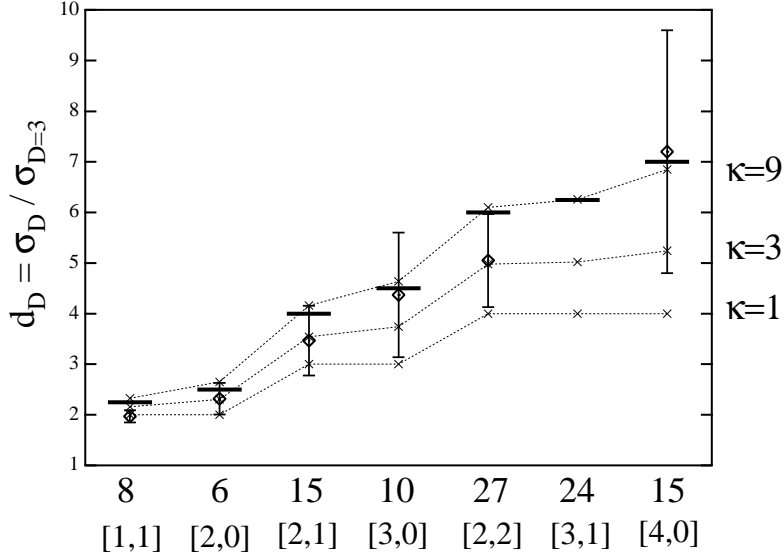}
\vspace*{-1cm}
\caption{\small 
The ratios of the string tensions of flux tubes for various
SU(3) representations, $d_{D}=\sigma_{D}/\sigma_{\bf 3}$ 
for the GL parameters $\kappa=1$, $3$ and $9$ (represented by 
crosses, each case connected by lines to guide the eye).  
The ratios of eigenvalues of the quadratic Casimir operators
are shown as black bars. 
For comparison the lattice data of 
Ref.~\cite{Deldar:1999vi} are also
plotted (diamonds with error bars).
Boldface numbers and brackets $[p,q]$ denote the
dimension and the Dynkin indices of each 
representation $D$, respectively}
\label{fig:casimir-abelian}
 \end{center}
 \vspace*{-0.6cm}
 \end{figure}

\par
In Fig.~\ref{fig:casimir-abelian}, we show the ratios of 
the string tensions of the flux tubes,
$d_{D}=\sigma_{D}/\sigma_{\bf 3}$  for three values
of the GL parameter, $\kappa = 1$, $3$, and $9$
(numerically obtained for $\kappa \ne 1$).
We also plot  the ratios of the string tensions obtained 
by the lattice simulations of Ref.~\cite{Deldar:1999vi}
and the ratios of eigenvalues of the quadratic Casimir 
operator,
\bea
 C^{(2)}(D) =
\frac{1}{3}(p^2+pq+q^2)+(p+q) .
\label{eqn:casimir}
\eea
We find that the DGL result in the type~II dual
superconducting vacuum near $\kappa = 3$ agrees 
well with all lattice data obtained in 
Ref.~\cite{Deldar:1999vi}, albeit with big errors.
The mechanism of the $\kappa$ dependence is 
understood as follows.  
In the Bogomol'nyi limit, $\kappa =1$, the ratio between
the string tensions of a higher and the fundamental 
representation satisfies the {\it flux counting rule}:
the string tension $\sigma_{D}$ is simply proportional
to the number of the color-electric Dirac strings inside 
the flux tube, as seen from Eq.~\eqref{eqn:st-exact}.  
With increasing $\kappa$, the interaction ranges of 
these fields get out of balance, and an excess of energy
appears because of the interaction between fundamental
flux tubes.
This leads to systematic deviations
from the counting rule.
Note that the deviation of $d_{D}$ from the counting 
rule grows toward higher representations $D$, 
since the number of fundamental flux which coexist 
in the flux tube of representation $D$ increases 
as the sum $p+q$ of Dynkin indices.
On the other hand, we also find that the DGL result 
at $\kappa = 9$, for the deeply type~II 
vacuum,  uniformly reproduces 
{\it Casimir-like} ratios, through the deviations 
from the flux counting rule.
In this analysis the higher 
dimensional flux tube in type~II vacuum is assumed 
to be stable against splitting into fundamental ones. 
In principle, there must be a certain minimal 
$q$-$\bar{q}$ distance where such a effect is 
not negligible, depending on the GL parameter.
However, in any case, the string tensions are
saturated by the values at the Bogomol'nyi limit
even if the splitting takes place.


\vspace*{-0.3cm}
\section{Summary} 
\vspace*{-0.3cm}

We have studied the string tensions of flux tubes
associated with static charges in various SU(3) 
representations in the DGL theory, based on a manifestly 
Weyl symmetric procedure.
We have found that a  GL-parameter near $\kappa = 3$
reproduces the ratios of string tensions
consistent with the lattice 
data~\cite{Deldar:1999vi}. 
The DGL theory accidentally shows Casimir-like scaling for 
a deeply type~II vacuum with $\kappa = 9$. 
However, there is no direct relation to the eigenvalues
of the Casimir operator.
The mechanism of the systematic behavior of string
tensions in the DGL theory can be understood as a 
result of the flux-tube dynamics. 
At present,  it is  not obvious that lattice
data really contain such a dynamical effect.
However, this example suggests that
it is important to have more lattice results 
carefully interpreted without bias toward the 
Casimir scaling hypothesis.


\vspace*{-0.2cm}
\section*{Acknowledgement}
\vspace*{-0.3cm}
We are grateful to E.-M. Ilgenfritz and T. Suzuki, 
for useful discussions.
This work is partially supported  by 
the Ministry of Education, Science, Sports and Culture,
Japan, Grant-in-Aid for Encouragement of Young 
Scientists (B), 14740161, 2002.


\end{document}